# CHSalign: A Web Server That Builds upon Junction-Explorer and RNAJAG for Pairwise Alignment of RNA Secondary Structures with Coaxial Helical Stacking

Lei Hua[1,¶], Yang Song[1,¶], Namhee Kim[2], Christian Laing[1], Jason T. L. Wang[1,*]

Tamar Schlick[2,3,*]

[1]Bioinformatics Laboratory, Department of Computer Science, New Jersey Institute of Technology, Newark, New Jersey, United States of America, [2]Department of Chemistry, [3]Courant Institute of Mathematical Sciences, New York University, New York, United States of America

[*]Corresponding authors

E-mail: wangj@njit.edu (JW), schlick@nyu.edu (TS)

[¶]These authors contributed equally to this work.

# Abstract

RNA junctions are important structural elements of RNA molecules. They are formed when three or more helices come together in three-dimensional space. Recent studies have focused on the annotation and prediction of coaxial helical stacking (CHS) motifs within junctions. Here we exploit such predictions to develop an efficient alignment tool to handle RNA secondary structures with CHS motifs. Specifically, we build upon our Junction-Explorer software for predicting coaxial stacking and RNAJAG for modelling junction topologies as tree graphs to incorporate constrained tree matching and dynamic programming algorithms into a new method, called CHSalign, for aligning the secondary structures of RNA molecules containing CHS motifs. Thus, CHSalign is intended to be an efficient alignment tool for RNAs containing similar junctions. Experimental results based on thousands of alignments demonstrate that CHSalign can align two RNA secondary structures containing CHS motifs more accurately than other RNA secondary structure alignment tools. CHSalign yields a high score when aligning two RNA secondary structures with similar CHS motifs or helical arrangement patterns, and a low score otherwise. This new method has been implemented in a web server, and the program is also made freely available, at http://bioinformatics.njit.edu/CHSalign/.

# Introduction

RNA secondary structures are composed of double-stranded segments such as helices connected to single-stranded regions such as junctions and hairpin loops. These structural elements serve as building blocks in the design of diverse RNA molecules with various functions in the cell [1-3]. In particular, RNA junctions are important structural elements due to their ability to orient many parts of the RNA molecule [4].

An RNA junction, also known as a multi-branch loop, forms when more than two helical segments are brought together [5-10]. RNA junctions exist in numerous RNA molecules; they play important roles in a wide variety of biochemical activities such as self-cleavage of the hammerhead ribozyme [11], the recognition of the binding pocket domain by purine riboswitches [12] and the translation initiation of the hepatitis C virus at the internal ribosome entry site [13]. Recent studies have classified RNA junctions with three and four branches into three and nine families, respectively [14,15]. Experiments have verified that a three-way junction in *Arabidopsis* has an important functional role [16]. A junction database, called RNAJunction, has been established, which contains junctions of all known degrees of branching [5].

A common tertiary motif within junctions of an RNA molecule is the coaxial stacking of helices [17-19], which occurs when two separate helical segments are aligned on a common axis to form a pseudocontiguous helix [20]. Coaxial stacking configurations have been observed in all large RNAs for which crystal structures are available, including tRNA, group I and II introns, RNase P, riboswitches and large ribosomal subunits. Coaxial helical stacking (CHS) provides thermodynamic stability to the RNA molecule as a whole [21] and reduces the separation between loop regions within junctions [22]. Moreover, coaxial stacking configurations form cooperatively with long-range interactions in many RNAs [14,17,23], and are therefore crucial as for correct tertiary structure formation as well as the formation of different junction topologies [15,17,24]. Since junctions are major architectural components in RNA, it is important to understand their structural properties. For example, the function of RNA molecules may be inferred if their junction components are similar in structure to other well-studied junction domains.

In this paper we build upon our previously developed Junction-Explorer tool [25] for predicting coaxial stacking and RNAJAG [4] for modelling junction topologies as tree graphs, and present a method, CHSalign, for aligning two RNA secondary (2D) structures that possess CHS motifs within the junctions of the two RNA structures. Coaxial stacking interactions in junctions are part of tertiary (3D) motifs [24]. Thus, CHSalign differs from both RNA 2D and 3D structure alignment tools. Existing secondary (2D) structure alignment tools focus on sequences and base pairs without considering tertiary motifs. Existing tertiary (3D) structure alignment tools accept as input two RNA 3D structures including all types of tertiary motifs in the Protein Data Bank (PDB) [26] and align the 3D structures by considering their geometric properties, torsion angles, and base pairs.

For 3D structure alignment, Ferre *et al.* [27] developed a dynamic programming algorithm based on nucleotide, dihedral angle, and base pairing similarities. Capriotti and Marti-Renom [28] developed a program to align two RNA 3D structures based on a unit-vector root-mean-square approach. Chang *et al.* [29] and Wang *et al.* [30] employed a structural alphabet of different nucleotide conformations to align RNA 3D structures. Hoksza and Svozil [31] developed a pairwise comparison method based on 3D similarity of generalized secondary structure units. Sarver *et al.* [32] designed the FR3D tool for finding local and composite recurrent structural motifs in RNA 3D structures. Dror *et al.* [33] described the RNA 3D structure alignment program, ARTS, and its use in the analysis and classification of RNA 3D structures [34]. Rahrig *et al.* [35] presented the R3D Align tool for performing global pairwise alignment of RNA 3D structures using local superpositions. He *et al.* [36] developed the RASS web server for comparing RNA 3D structures using both sequence and 3D structure information.

On the other hand, a well-adopted strategy for RNA 2D structure alignment is to use a tree transformation technique and perform RNA alignment through tree matching [37-39]. For instance, RNAforester [39] aligns two RNA 2D structures by calculating the edit-distance between tree structures symbolizing RNAs. By utilizing tree models to capture the structural particularities in RNA, RSmatch [37] aligns two RNA 2D structures effectively. Additional methods are described in [38,40].

In contrast to these methods for aligning two RNAs when their 2D structures are available, another group of closely related methods achieved RNA folding and alignment simultaneously. For instance, FOLDALIGN [41] uses a lightweight energy model and sequence similarity to simultaneously fold and align RNA sequences. Dynalign [42] finds a secondary structure common to two sequences without requiring any sequence identity. DAFS [43] simultaneously aligns and folds RNA sequences based on maximizing the expected accuracy of a predicted common secondary structure of the sequences. Similar techniques are implemented in CentroidAlign [44] and SimulFold [45]. SCARNA [46] employs a method of comparing RNA sequences based on the structural alignment of the fixed-length fragments of the stem candidates in the RNAs.

While many methods have been developed for RNA structure alignment, as surveyed above, few are tailored to junctions, especially junctions with coaxial stacking interactions. Junctions and coaxial stacking patterns are common in many RNA molecules and, as mentioned above, are involved in a wide range of functions. Furthermore, experimental probing techniques, such as RNA SHAPE chemistry, SAXS, NMR, and fluorescence resonance energy transfer (FRET), often provide sufficient information to determine coaxial stacking configurations [2,47-49]. Thus, a junction-tailored tool capable of comparing RNA structures on the basis of coaxial stacking patterns in their junctions could be particularly valuable. To this end, we present CHSalign, which performs RNA alignment by applying a constrained tree matching algorithm and dynamic programming techniques to ordered labeled trees symbolizing RNA structures with coaxial stacking patterns. Experimental results on different data sets demonstrate the effectiveness of this newly developed tool. The CHSalign web server is freely available at http://bioinformatics.njit.edu/CHSalign/.

## Materials and Methods

CHSalign accepts as input two RNA 2D structures which contain manually annotated coaxial stacking of helices, and produces as output an alignment between the two input structures. When manually annotated coaxial stacking patterns are not available, CHSalign invokes our previously developed Junction-Explorer tool [25] to predict the coaxial stacking configurations of the input structures.

   Our approach is to transform each input RNA 2D structure with coaxial stacking patterns into an ordered labeled tree. Tree graphs are popular models for representing RNA structures [4,23,39,50-52]. We extend RNAJAG [4] to obtain an ordered tree model, in which each tree node represents a secondary structure element such as a helix (stem), junction or hairpin loop. When comparing two tree nodes, we use a dynamic programming algorithm [37,38] to align the 2D structural elements in the tree nodes, obtaining a score between the two nodes. We then use a constrained tree matching algorithm to find an optimal alignment between the two input RNA 2D structures, taking into account their coaxial stacking configurations. Below, we detail our tree model and the constrained tree matching algorithm.

**Tree model formalism**

Let $R_{seq}$ be an RNA sequence containing nucleotides or bases A, C, G, U. $R_{seq}[i]$ denotes the base at position $i$ of $R_{seq}$ ordered from the 5′ to 3′ ends. $R_{seq}[i, j]$, $i < j$, is the subsequence starting at position $i$ and ending at position $j$. Let $R$ be the 2D structure of $R_{seq}$ with at least one base pair. A helix in $R$ is a double-stranded segment composed of contiguous base pairs. A base pair connecting position $i$ and position $j$ is denoted by $(i, j)$ and its enclosed subsequence is $R_{seq}[i, j]$. If all nucleotides in $R_{seq}[i, j]$ except $R_{seq}[i]$ and $R_{seq}[j]$ are unpaired single bases, and $(i, j)$ is a base pair in $R$, we call $R_{seq}[i+1, j-1]$ a hairpin loop.

A junction, or a multi-branch loop, is an enclosed area connecting different helices [7]. An $n$-way junction in $R$ has $n$ branches. This junction connects $n$ helices where there are $n$ base pairs $(i_1, j_1)... (i_n, j_n)$ (one base pair for each helix), and $n$ subsequences participating in the junction. The $n$ subsequences are denoted by $R_{seq}[i_1+1, i_2-1]$, $R_{seq}[j_2+1, i_3-1]$, $R_{seq}[j_3+1, i_4-1]$, ... , $R_{seq}[j_{n-1}+1, i_n-1]$, and $R_{seq}[j_n+1, j_1-1]$. All the unpaired bases on the $n$ subsequences comprise the $n$-way junction, and the subsequences are called the loop regions of the junction. Internal loops or bulges can be considered as special cases of "two-way" junctions [5]. However, for the purpose of this work, $n$ must be greater than 2. Thus, internal loops or bulges are not considered as junctions in our work; instead, they are considered as part of the helices in $R$.

We transform the 2D structure $R$ into an ordered labeled tree $T$ in which each node has a label and the left-to-right order among sibling nodes is important. Each node of $T$ represents a 2D structural element of *R*, belonging to one of three types: helix, junction, and hairpin loop. With this tree model, pseudoknots are excluded.

Figure 1 illustrates the transformation process. Figure 1(A) shows the 3D crystal structure of the adenine riboswitch molecule (PDB code: 1Y26) obtained from the Protein Data Bank (PDB) [26] and drawn by PyMOL (http://www.pymol.org/). The first helix according to the 5′ to 3′ orientation is labeled by $H_1$ and highlighted in blue. The second helix is labeled by $H_2$ and highlighted in green. The third helix is labeled by $H_3$ and highlighted in red. The junction labeled by $J_1$ and hairpin loops labeled by $P_1$ and $P_2$ respectively are highlighted in light grey. $J_1$ is a multi-branch loop where the three helices $H_1$, $H_2$ and $H_3$ connect. $P_1$ and $P_2$ are hairpin loops connected to helices $H_2$ and $H_3$, respectively.

Figure 1(B) shows the corresponding 2D structure, obtained from RNAView [53]. Each 2D structural element in Figure 1(B) is highlighted as in Figure 1(A). Notice that there is a yellow bar across $H_1$, $J_1$ and $H_3$, symbolizing a coaxial helical stacking $H_1H_3$ in the molecule 1Y26, as described in [17,24]. In general, the coaxial helical stacking status of a three-way junction such as $J_1$ in Figure 1(B) is described as one of four possibilities: $H_1H_2$, $H_2H_3$, $H_1H_3$, or none, where $H_xH_y$ indicates that $H_x$ and $H_y$ are coaxially stacked, i.e., helix $H_x$ shares a common axis with helix $H_y$. The locations of the junctions and the coaxial helical stacking status of each junction in a given 2D structure can be determined using the methods described in [25].

Figure 1(C) shows the tree, *T*, used to represent the 2D structure *R* in Figure 1(B). Each node of *T* corresponds to a 2D structural element of *R* where the octagon (squares, triangles respectively) in *T* represents the junction (helices, hairpin loops respectively) in *R*. Thus, like the 2D structural elements, each tree node belongs to one of three types, namely helix, junction, and hairpin loop. Tree nodes of different types are prohibited to be aligned with each other, and hence the term "constrained tree matching" is used in our work (reminiscent of structural constraints in RNA described in [54]).

We use $t[i]$ to represent the node of tree $T$ whose position in the left-to-right post-order traversal of $T$ is $i$. The post-order procedure works by first traversing the left subtree, then traversing the right subtree, and finally visiting the root. In Figure 1(C), the post-order position number of each node is shown next to the node. By construction, the tree node corresponding to an $n$-way junction consists of $n - 1$ children. The first helix according to the 5′ to 3′ orientation is the parent node of the junction node. The other $n - 1$ helices are the children of that junction node. The number of children of node $t[i]$ is the degree of $t[i]$. In Figure 1(C), $H_1$ is the parent node of $J_1$, which has two children, $H_2$ and $H_3$. The degree of the junction node $J_1$ is 2. In general, the degree of an $n$-way junction node is $n - 1$.

Consider two RNA 2D structures $R_1$ and $R_2$ and their tree representations $T_1$ and $T_2$ respectively. Let $t_1[i]$ ($t_2[j]$, respectively) be the node of $T_1$ ($T_2$, respectively) whose position in the post-order traversal of $T_1$ ($T_2$, respectively) is $i$ ($j$, respectively). Let $T_1[i]$ be the subtree rooted at $t_1[i]$, and $T_2[j]$ be the subtree rooted at $t_2[j]$. $F_1[i]$ represents the forest obtained by removing the root $t_1[i]$ from subtree $T_1[i]$. $F_2[j]$ represents the forest obtained by removing the root $t_2[j]$ from subtree $T_2[j]$. Suppose the degree of $t_1[i]$ is $m_i$ (i.e., $t_1[i]$ has $m_i$ children $t_1[i_1], \ldots, t_1[i_{m_i}]$) and the degree of $t_2[j]$ is $n_j$ (i.e., $t_2[j]$ has $n_j$ children $t_2[j_1], \ldots, t_2[j_{n_j}]$). We use $S(T_1[i], T_2[j])$ to represent the alignment score of subtree $T_1[i]$ and subtree $T_2[j]$, and use $\gamma(t_1[i], t_2[j])$ to represent the alignment score of node $t_1[i]$ and node $t_2[j]$. We use $\emptyset$ to represent an empty node; matching a tree node with $\emptyset$ amounts to aligning all nucleotides in the tree node to gaps.

**Alignment scheme**

We employ a dynamic programming algorithm to align two RNA 2D structures with coaxial stacking patterns. Our approach is to transform each RNA 2D structure into an ordered labeled tree as explained in the previous subsection. We then apply the dynamic programming algorithm to the ordered labeled trees representing the two RNA 2D structures. Based on the alignment of the trees, we obtain the alignment of the corresponding RNA 2D structures. As noted above, each tree node belongs to one of three types: helix, junction, and hairpin loop. Different types of tree nodes are prohibited to be aligned with each other. When aligning two subtrees $T_1[i]$ and $T_2[j]$ and calculating the score $S(T_1[i], T_2[j])$, there are nine cases to be considered.

**Case 1.** Both $t_1[i]$ and $t_2[j]$ are junctions.

One constraint we impose on pairwise alignment is that when aligning a $p$-way junction node $v_1$ with a $q$-way junction node $v_2$, $p$ must be equal to $q$. Furthermore, the coaxial helical stacking status of $v_1$ must be the same as the coaxial stacking status of $v_2$. Thus, a three-way junction must be aligned with a three-way junction, which is not allowed to align with a four-way junction. Furthermore, a three-way junction whose coaxial helical stacking status is $H_1H_2$ must be aligned with a three-way junction having the same $H_1H_2$ status, which is not allowed to align with a three-way junction whose coaxial helical stacking status is $H_2H_3$. In general, junctions with different branches and different coaxial stacking configurations have different biological properties. This constraint is established to ensure a biologically meaningful alignment is obtained, and to avoid introducing too many gaps in the alignment.

According to our tree model, if a tree node is a junction, it must have at least two children and the children must be helix nodes. A junction contains loop regions with single bases whereas helices are double-stranded regions with base pairs. A junction node is thus prohibited to be aligned with a helix node. Hence, $t_1[i]$ must be aligned with $t_2[j]$ provided they have the same number of branches and the same coaxial helical stacking status, denoted by $\Psi(t_1[i]) = \Psi(t_2[j])$. Their children are trees, which together form forests $F_1[i]$ and $F_2[j]$ respectively. $F_1[i]$ must be aligned with $F_2[j]$. Thus the alignment score of $T_1[i]$ and $T_2[j]$ can be calculated as:

$$S(T_1[i], T_2[j]) = \max \begin{cases} \gamma(t_1[i], t_2[j]) + S(F_1[i], F_2[j]) \\ 0 \end{cases}. \quad (1)$$

If $\Psi(t_1[i]) = \Psi(t_2[j])$, $t_1[i]$ and $t_2[j]$ must have the same number of children, and the order among the sibling nodes is important. If $\Psi(t_1[i]) \neq \Psi(t_2[j])$, i.e., $t_1[i]$ and $t_2[j]$ have different numbers of children (branches) or they have different coaxial helical stacking statuses, they are prohibited to be aligned together. Thus, the score of matching $F_1[i]$ with $F_2[j]$ can be calculated as:

$$S(F_1[i], F_2[j]) = \begin{cases} S(T_1[i_1], T_2[j_1]) + \ldots + S(T_1[i_m], T_2[j_m]) & \text{if } \Psi(t_1[i]) = \Psi(t_2[j]) \\ -\infty & \text{otherwise} \end{cases} \quad (2)$$

where $m$ is the number of children of $t_1[i]$ and $t_2[j]$ respectively. We use $\Pi(t_1[i])$ to represent the coaxial helical stacking status of $t_1[i]$; $\Pi(t_1[i]) = 1$ (2, 3, 0, respectively) if the coaxial helical stacking status of $t_1[i]$ is $H_1H_2$ ($H_2H_3$, $H_1H_3$, none, respectively). The score of matching $t_1[i]$ with $t_2[j]$ is

$$\gamma(t_1[i], t_2[j]) = \begin{cases} s+w & \text{if } \Psi(t_1[i]) = \Psi(t_2[j]), \Pi(t_1[i]) \neq 0, \Pi(t_2[j]) \neq 0 \\ s+w/2 & \text{if } \Psi(t_1[i]) = \Psi(t_2[j]), \Pi(t_1[i]) = \Pi(t_2[j]) = 0 \\ -\infty & \text{otherwise} \end{cases} \quad (3)$$

Here, $s$ is the score obtained by aligning the junction in $t_1[i]$ with the junction in $t_2[j]$. We use a dynamic programming algorithm [37,38] to calculate the alignment score $s$, and adopt the RIBOSUM85-60 matrix [55] to calculate the score of aligning two bases or base pairs in RNA 2D structures. (The default gap penalty is $-1$.) With this scoring matrix, CHSalign can handle non-canonical base pairs. The addition of a parameter $w$ to the alignment score is a computational device to enforce the right alignment of the RNAs when the junction patterns match. Thus, if $t_1[i]$ and $t_2[j]$ have the same number of branches, their CHS patterns are alike, and $\Pi(t_1[i]) \neq 0$, $\Pi(t_2[j]) \neq 0$, we use $s+w$ as the modified alignment score. When $t_1[i]$ and $t_2[j]$ have the same number of branches and $\Pi(t_1[i]) = \Pi(t_2[j]) = 0$, we use $s+(w/2)$ as the modified score. The value of $w$ required experimentation, as we discuss later, but a value of 100 seems to work well in practice.

**Case 2.** Both $t_1[i]$ and $t_2[j]$ are helices.

Due to the nature of RNA 2D structures and based on our tree model, a helix has only one child, which is either a junction or a hairpin loop. The subtree rooted at the child of $t_1[i]$ is denoted by $T_1[i-1]$ and the subtree rooted at the child of $t_2[j]$ is denoted by $T_2[j-1]$. We have to match helix nodes $t_1[i]$ and $t_2[j]$ first, and then add the alignment score of their subtrees $T_1[i-1]$ and $T_2[j-1]$ if the alignment score of the subtrees is greater than or equal to zero, or simply match $t_1[i]$ with $t_2[j]$ if the alignment score of their subtrees is negative (i.e., the subtrees are not aligned). Therefore, the alignment score of $T_1[i]$ and $T_2[j]$ can be calculated as:

$$S(T_1[i], T_2[j]) = \max \begin{cases} \gamma(t_1[i], t_2[j]) + S(T_1[i-1], T_2[j-1]) \\ \gamma(t_1[i], t_2[j]) \\ 0 \end{cases} \qquad (4)$$

The score $\gamma(t_1[i], t_2[j])$ is obtained by aligning the helix in $t_1[i]$ with the helix in $t_2[j]$ using a dynamic programming algorithm [37,38]. The value 0 is used if the other entries in Equation (4) yield negative scores.

**Case 3.** Both $t_1[i]$ and $t_2[j]$ are hairpin loops.

Due to the nature of RNA 2D structures and based on our tree model, a hairpin does not have any child. Therefore hairpin nodes are always leaves in the tree representation of an RNA 2D structure. When both $t_1[i]$ and $t_2[j]$ are hairpin loops, matching $T_1[i]$ with $T_2[j]$ amounts to matching $t_1[i]$ with $t_2[j]$. Thus, the alignment score becomes:

$$S(T_1[i], T_2[j]) = \max \begin{cases} \gamma(t_1[i], t_2[j]) \\ 0 \end{cases} \qquad (5)$$

The score $\gamma(t_1[i], t_2[j])$ is obtained by aligning the hairpin loop in $t_1[i]$ with the hairpin loop in $t_2[j]$ using a dynamic programming algorithm [37,38].

**Case 4.** $t_1[i]$ is a junction and $t_2[j]$ is a helix.

Since $t_1[i]$ and $t_2[j]$ have different types, they cannot be aligned with each other. There are two subcases.

Subcase 1. $t_2[j]$ is aligned to gaps. Then $T_1[i]$ must be aligned with $T_2[j-1]$, which is the subtree rooted at the child of $t_2[j]$.

Subcase 2. $t_1[i]$ is aligned to gaps. Suppose $t_1[i]$ has $m_i$ children $t_1[i_1], \ldots, t_1[i_{m_i}]$. The subtrees rooted at these children are denoted by $T_1[i_1], \ldots, T_1[i_{m_i}]$ respectively. Then, one of these subtrees must be aligned with $T_2[j]$; specifically the subtree yielding the maximum alignment score is aligned with $T_2[j]$.

We take the maximum of the above two subcases. Thus, the score of matching $T_1[i]$ with $T_2[j]$ can be calculated as:

$$S(T_1[i], T_2[j]) = \max \begin{cases} \gamma(\varnothing, t_2[j]) + S(T_1[i], T_2[j-1]) \\ \gamma(t_1[i], \varnothing) + \max_{1 \leq k \leq m_i} \{S(T_1[i_k], T_2[j])\} \\ 0 \end{cases}. \quad (6)$$

The value 0 is used if both of the two subcases yield negative scores.

Figure 2 illustrates this case where two PDB molecules, A-riboswitch (PDB code: 1Y26) and the Alu domain of the mammalian signal recognition particle (SRP) (PDB code: 1E8O), are considered. Figure 2(A) shows the 3D crystal structure of the adenine riboswitch molecule and its tree representation $T_1$. Figure 2(B) shows the 3D crystal structure of the Alu domain of the mammalian SRP molecule and its tree representation $T_2$. When matching $T_1[i]$ with $T_2[j]$, since $t_1[i]$ and $t_2[j]$ have different types where $t_1[i]$ is a junction and $t_2[j]$ is a helix, there are two subcases to be considered, as detailed above. Figure 2(C-i) illustrates subcase 1, in which $t_2[j]$ is aligned to gaps and $T_1[i]$ is aligned with $T_2[j-1]$. Figure 2(C-ii) illustrates subcase 2, in which $t_1[i]$ is aligned to gaps, and the subtree rooted at one of the children of $t_1[i]$ is aligned with $T_2[j]$. In our example here, $t_1[i]$ has two children, $t_1[i_1]$ and $t_1[i_2]$. Thus, either the subtree rooted at $t_1[i_1]$, denoted by $T_1[i_1]$, is aligned with $T_2[j]$ as illustrated in Figure 2(C-iia), or the subtree rooted at $t_1[i_2]$, denoted by $T_1[i_2]$, is aligned with $T_2[j]$ as illustrated in Figure 2(C-iib). The maximum alignment score obtained from Figure 2(C-iia) and Figure 2(C-iib) is used. Then $S(T_1[i], T_2[j])$ is calculated by taking the maximum of the two subcases illustrated in Figure 2(C-i) and Figure 2(C-ii) respectively.

**Case 5.** $t_1[i]$ is a junction and $t_2[j]$ is a hairpin loop.

Since $t_1[i]$ and $t_2[j]$ have different types, the two nodes cannot be aligned together. Furthermore, $t_2[j]$ is a hairpin loop, which does not have any child. Thus $t_1[i]$ must be aligned to gaps, and the subtree rooted at one of the children of $t_1[i]$ is aligned with $T_2[j]$; specifically the subtree yielding the maximum alignment score is aligned with $T_2[j]$. Therefore, the alignment score of $T_1[i]$ and $T_2[j]$ can be calculated as:

$$S(T_1[i], T_2[j]) = \max \begin{cases} \gamma(t_1[i], \emptyset) + \max_{1 \leq k \leq m_i} \{S(T_1[i_k], T_2[j])\} \\ 0 \end{cases}. \quad (7)$$

**Case 6.** $t_1[i]$ is a helix and $t_2[j]$ is a junction.

Similar to Case 4, there are two subcases.

Subcase 1. $t_i[i]$ is aligned to gaps. Thus, the subtree rooted at the child of $t_1[i]$, denoted by $T_1[i-1]$, must be aligned with $T_2[j]$.

Subcase 2. $t_2[j]$ is aligned to gaps. Suppose $t_2[j]$ has $n_j$ children $t_2[j_1], \ldots, t_2[j_{n_j}]$. The subtrees rooted at these children are $T_2[j_1], \ldots, T_2[j_{n_j}]$ respectively. Then $T_1[i]$ must be aligned with one of these subtrees.

Taking the maximum of these two subcases, we calculate the score of matching $T_1[i]$ with $T_2[j]$ as:

$$S(T_1[i], T_2[j]) = \max \begin{cases} \gamma(t_1[i], \emptyset) + S(T_1[i-1], T_2[j]) \\ \gamma(\emptyset, t_2[j]) + \max_{1 \leq k \leq n_j} \{S(T_1[i], T_2[j_k])\} \\ 0 \end{cases}. \quad (8)$$

**Case 7.** $t_1[i]$ is a helix and $t_2[j]$ is a hairpin loop.

Because $t_1[i]$ and $t_2[j]$ have different types, the two nodes cannot be aligned together. Furthermore, since $t_1[i]$ is a helix, it has only one child; $t_2[j]$ is a hairpin loop with no children. Therefore, $t_1[i]$ must be aligned to gaps and the subtree rooted at the child of $t_1[i]$, denoted by $T_1[i-1]$, must be aligned with $T_2[j]$, or if the alignment yields a negative score, we use the value 0. Thus, the alignment score is

$$S(T_1[i], T_2[j]) = \max \begin{cases} \gamma(t_1[i], \varnothing) + S(T_1[i-1], T_2[j]) \\ 0 \end{cases}. \tag{9}$$

**Case 8.** $t_1[i]$ is a hairpin loop and $t_2[j]$ is a junction.

This is similar to Case 5. Thus, we can calculate the score of matching $T_1[i]$ with $T_2[j]$ as:

$$S(T_1[i], T_2[j]) = \max \begin{cases} \gamma(\varnothing, t_2[j]) + \max_{1 \le k \le n_j} \{ S(T_1[i], T_2[j_k]) \} \\ 0 \end{cases}. \tag{10}$$

**Case 9.** $t_1[i]$ is a hairpin loop and $t_2[j]$ is a helix.

This is similar to Case 7, with the alignment score:

$$S(T_1[i], T_2[j]) = \max \begin{cases} \gamma(\varnothing, t_2[j]) + S(T_1[i], T_2[j-1]) \\ 0 \end{cases}. \tag{11}$$

**Time and space complexity**

Let $|T_1|$ ($|T_2|$ respectively) denote the number of nodes in tree $T_1$ ($T_2$ respectively) that represents RNA structure $R_1$ ($R_2$ respectively). CHSalign maintains a two-dimensional table in which $c(i, j)$ represents the cell located at the intersection of the $i$th row and the $j$th column of the table. The value stored in the cell $c(i, j)$, $1 \le i \le |T_1|$, $1 \le j \le |T_2|$, is $S(T_1[i], T_2[j])$. The dynamic programming algorithm employed by CHSalign calculates the values in the table by traversing the trees $T_1$ and $T_2$ in a bottom-up manner. After all the values in the table are computed, the algorithm locates the cell $c$ with the maximum value. A backtrack procedure starting with the cell $c$ and terminating when encountering a zero identifies the alignment lines of an optimal alignment and calculates the alignment score between $T_1$ and $T_2$.

Let $|R_1|$ ($|R_2|$ respectively) denote the number of nucleotides, i.e., the length, of RNA structure $R_1$ ($R_2$ respectively). Let $|t_1[i]|$ ($|t_2[j]|$ respectively) be the number of nucleotides in node $t_1[i]$ ($t_2[j]$ respectively). Let $d_1$ ($d_2$, respectively) be the maximum degree of any node in tree $T_1$ ($T_2$ respectively). The time complexity of computing $\gamma(t_1[i], t_2[j])$ is $O(|t_1[i]| \times |t_2[j]|)$ [37]. Thus, the time complexity of computing $S(T_1[i], T_2[j])$ is $O(\max(d_1, d_2) + |t_1[i]| \times |t_2[j]|)$. Here $\max(d_1, d_2)$ is a constant because a junction has at most twelve branches in solved RNA crystal structures [4,25,52]. Furthermore, $\sum_{i=1}^{|T_1|} |t_1(i)| = |R_1|$ and $\sum_{j=1}^{|T_2|} |t_2(j)| = |R_2|$. Therefore the time complexity of calculating all the values in the two-dimensional table is

$$\begin{aligned} &O\left(\sum_{i=1}^{|T_1|} \sum_{j=1}^{|T_2|} \left(\max(d_1, d_2) + |t_1[i]| \times |t_2[j]|\right)\right) \\ &= O\left(\sum_{i=1}^{|T_1|} \sum_{j=1}^{|T_2|} \left(|t_1[i]| \times |t_2[j]|\right)\right) \\ &= O(|R_1| \times |R_2|). \end{aligned} \quad (12)$$

Locating the cell $c$ with the maximum value in the two-dimensional table and executing the backtrack procedure require $O\left(\sum_{i=1}^{|T_1|} \sum_{j=1}^{|T_2|} (|t_1[i]| \times |t_2[j]|)\right) = O(|R_1| \times |R_2|)$ computational time. Therefore the time complexity of CHSalign is $O(|R_1| \times |R_2|)$. Since only a two-dimensional table is used, the space complexity of CHSalign is $O(|T_1| \times |T_2|)$ = $O(|R_1| \times |R_2|)$.

**Data sets**

Popular benchmark datasets such as BRAliBase [56] and Rfam [57] are not suitable for testing CHSalign, since they do not contain coaxial helical stacking information. As a consequence, we manually created two datasets for testing CHSalign and comparing it with related methods. The first dataset, Dataset1, contains 24 RNA 3D structures from the Protein Data Bank (PDB) [26] (see Table 1). This dataset was studied and published in [4,25,52], in which all annotations for junctions and coaxial helical stacking were taken from crystallographic structures. Each 3D structure in Dataset1 contains at least one three-way junction, and the lengths of the 3D structures range from 40 nt to 2,958 nt. Some 3D structures contain higher-order junctions such as ten-way junctions with coaxial stacking patterns. The 2D structure of each 3D structure in Dataset1 is obtained with RNAView retrieved from RNA STRAND [58]. The pseudoknots in these structures are removed using the K2N tool [59].

The second dataset, Dataset2, contains 76 three-way junctions extracted from the 24 3D structures in Dataset1. (Some 3D structures in Dataset1 contain more than one three-way junction and all those three-way junctions in a 3D structure are extracted.) The lengths of the three-way junctions range from 28nt to 153nt. The coaxial helical stacking status of each three-way junction in Dataset2 is described as one of three possibilities: $H_1H_2$, $H_2H_3$, $H_1H_3$. Thus, every three-way junction in Dataset2 contains a coaxial stacking pattern. In the RNA literature, most research efforts have been focused on three-way and four-way junctions [6,15,60-62] partly due to the fact that higher-order junctions are rare. In particular, three-way junctions are the most abundant type of junctions, accounting for over 50% of the available crystal data. We also performed experiments on four-way junctions; results obtained from the four-way junctions were similar to those for the three-way junctions reported here, and hence omitted.

## Results and Discussion

### Two CHSalign web server versions

We have implemented two programs in Java, a standalone version denoted by CHSalign_u, and the other a pipeline denoted by CHSalign_p. CHSalign_u requires the user to manually annotate the coaxial stacking patterns within junctions of the pair of RNA 2D structures in the input, and produces an optimal alignment between the two input structures.

By contrast, CHSalign_p accepts as input two unannotated RNA 2D structures and produces as output an optimal alignment between the two input structures while taking into account their junctions and coaxial stacking configurations within the junctions. This pipeline invokes our previously developed Junction-Explorer tool [25] to automatically predict and identify the junctions and coaxial stacking patterns within the junctions in the input structures, and then aligns the input structures containing the predicted coaxial stacking patterns. Both CHSalign_u and CHSalign_p are available on the web.

### Performance evaluation using RMSD

We conducted a series of experiments to evaluate the performance of our algorithms. In the first experiment, we divided Dataset2 into three disjoint subsets Dataset2-1, Dataset2-2 and Dataset2-3, with 35, 18, and 23 junctions, respectively. These three subsets contain, respectively, three-way junctions whose coaxial helical stacking status is $H_1H_2$, $H_2H_3$, or $H_1H_3$. We performed pairwise alignment of junctions in each subset. There are $(35 \times 34/2 + 18 \times 17/2 + 23 \times 22/2) = 1,001$ pairwise alignments produced by CHSalign. Commonly used ways for evaluating the accuracy of these structural alignments include the calculation of distance matrices or RMSD (root-mean-square deviation) [4,29,32,63-66]. We adopt the RMSD measure [4,29] to evaluate the performance of our algorithms; specifically we use the method for computing RMSDs of tree graphs [4]. It has been shown that RMSDs of tree graphs and RMSDs of atomic models are positively correlated and indicate similar trends [4]. The average of the RMSD values of the 1,001 pairwise alignments was calculated and plotted.

One important parameter in our algorithms is the weight $w$ used in Equation (3) for calculating the alignment score of two junction nodes. This parameter is introduced to favor the alignment between two junctions with the same number of branches and the same coaxial helical stacking status. Experimental results show that when $w$ is sufficiently large (e.g., $w > 50$), our algorithms work well. In subsequent experiments, we fixed the weight $w$ in Equation (3) at 100.

Figure 3 compares CHSalign_u and CHSalign_p with three other alignment programs: RNAforester [39], RSmatch [37] and FOLDALIGN [41]. Like CHSalign, both RNAforester and RSmatch produce an alignment between two input RNA 2D structures. FOLDALIGN differs from the other programs in Figure 3 in that it performs 2D structure prediction and alignment simultaneously. When running the FOLDALIGN tool, the structure information in the datasets was ignored and only the sequence data was used as the input of the tool. In addition, when experimenting with CHSalign_u, the coaxial stacking patterns were provided along with the input RNA 2D structures. When running the other programs including CHSalign_p, RNAforester, RSmatch and FOLDALIGN, these coaxial stacking patterns were absent in the input. CHSalign_p automatically predicts the coaxial stacking patterns and then aligns the predicted structures.

Figure 3 shows that CHSalign_u performs the best, achieving an RMSD of 1.78 Å. The drawback of CHSalign_u, however, is that it requires the user to annotate the input RNA structures with coaxial stacking patterns manually. Manually annotating coaxial stacking patterns on RNA structures requires domain related expertise. On the other hand, CHSalign_p does not require any manual processing and achieves a reasonably good RMSD of 1.83 Å. Since the predicted coaxial stacking patterns may be imperfect, the RMSD of CHSalign_p is larger than that of CHSalign_u. RSmatch and RNAforester have even larger RMSDs of 4.41 Å and 6.13 Å, respectively. This happens because RSmatch and RNAforester ignore coaxial stacking configurations when aligning RNA 2D structures. FOLDALIGN has the largest RMSD of 8.26 Å, partly because it does not consider coaxial helical stacking either, and partly because there are errors in its predicted 2D structures.

**Performance evaluation using precision**

In the next experiment, we adopt *precision* as the performance measure, defined below, to evaluate how junctions and coaxial stacking patterns are aligned by different programs using the 24 structures in Dataset1. We say a junction $J_1$ in structure $R_1$ is aligned with a junction $J_2$ in structure $R_2$, or more precisely there is a junction alignment between $J_1$ and $J_2$, if there exist a nucleotide $n_1$ on a loop region of $J_1$ and a nucleotide $n_2$ on a loop region of $J_2$ such that $n_1$ is aligned with $n_2$. A junction alignment between $J_1$ and $J_2$ is a true positive if $J_1$ and $J_2$ have the same number of branches and the same coaxial helical stacking status. A junction alignment between $J_1$ and $J_2$ is a false positive if $J_1$ and $J_2$ have different numbers of branches or different coaxial helical stacking statuses. The precision (PR) of an alignment between $R_1$ and $R_2$ is defined as

$$PR = TP/(TP + FP), \tag{13}$$

where *TP* equals the number of true positives and *FP* equals the number of false positives in the alignment. The higher PR value a program has, the more precise alignment that program produces. In the experiment, we also included a closely related RNA 3D alignment tool (SETTER) [31].

We calculated the precision of each alignment produced by a program, took the average of the precision values of the pairwise alignments of the 24 structures in Dataset1, and plotted the average values. Figure 4 shows the result. We see that CHSalign_u performs the best, achieving a PR value of 1. CHSalign_p achieves a PR value of 0.85, not 1, because some coaxial stacking patterns were not predicted correctly by Junction-Explorer [25] used in CHSalign_p. The other programs in Figure 4 did not consider coaxial helical stacking while performing pairwise alignments, and hence achieved low PR values. Specifically, the PR values of RNAforester, SETTER, RSmatch, and FOLDALIGN were 0.54, 0.42, 0.33, and 0.31 respectively. Unlike the CHSalign method, these programs occasionally align two junctions with different numbers of branches or different coaxial helical stacking statuses, hence yielding false positives. However, SETTER is a general-purpose structure alignment tool capable of comparing two RNA 3D molecules with diverse tertiary motifs, while CHSalign can only deal with the 2D structures of the 3D molecules that contain coaxial helical stacking motifs.

## Potential application of CHSalign

To demonstrate the utility of the CHSalign tool, we applied CHSalign to the analysis of riboswitches that regulate gene expression by selectively binding metabolites [67]. Table 2 lists six riboswitches that bind to different metabolites (purine, guanine, thiamine pyrophosphate [TPP], and S-Adenosyl methionine [SAM]) found in different organisms. Since such binding and gene regulation activities are correlated to junction structures, the results of junction alignments could help suggest structural similarity (and thus possibly function) of these riboswitches. For each riboswitch, Table 2 also lists the junction type and coaxial helical stacking status within the junction in that riboswitch. Figure 5 illustrates the coaxial stacking patterns in the six riboswitches. We tested several combinations of junctions in these six riboswitches to determine whether the CHSalign results confirm known structural and functional similarity in existing RNAs. Table 3 summarizes the test results. Details of these results, including the input and output of each test, can be found in S1 File and S2 File.

Without knowledge of junction helical arrangements, we first tested the following cases using CHSalign_p, where the two aligned junctions had the same coaxial stacking patterns. We used SAM riboswitches in different organisms (PDB codes 2GIS and 4B5R in Table 2) as input. CHSalign_p predicted that the two riboswitches had helical arrangements of four-way junctions both with coaxial stacking helices 1 and 4 and helices 2 and 3, and produced a very high alignment score of 252.61, as calculated by the equations in the subsection 'Alignment scheme" in the section 'Materials and Methods'. This high score implies that the two riboswitches have highly similar helical arrangements. This corroborates our expectations, because the two tested riboswitches have similar structures and functionality, binding to SAM. Next, when we used purine and guanine riboswitches (PDB codes 2G9C and 3RKF), we obtained a high alignment score of 179.68 for three-way junction alignment of the two riboswitches with predicted coaxial stacking of helices 1 and 3 in both riboswitches, indicating high similarities of their three-way junction structures. We also tested two TPP riboswitches with three-way junctions in different organisms (PDB codes 2GDI and 3D2G), which produced a high alignment score of 191.06, again indicating that these two TPP riboswitches have similar three-way junction structures.

We next compared very different junction structures using CHSalign_p. When we aligned two different riboswitches – SAM riboswitch with a four-way junction and purine riboswitch with a three-way junction (PDB codes 2GIS and 2G9C, respectively), we obtained a low alignment score of 20.40. We also tested a pair of purine and TPP riboswitches (PDB codes 2G9C and 2GDI), which are in different riboswitch classes and have different coaxial stacking patterns in their three-way junctions. We obtained a low alignment score of 13.65. These experiments suggest that CHSalign_p, based only on secondary structural information, is useful for inferring tertiary structural features regarding helical arrangements.

Finally, we tested CHSalign_u, which requires prior information about junction arrangement and produces a structural similarity score for two given RNAs. Here, we tested two cases. First, we considered the same RNA structure (purine riboswitch with PDB code 2G9C) but annotated it with different helical arrangement patterns where one had coaxial stacking helices 1 and 3 ($H_1H_3$) and the other had coaxial stacking helices 1 and 2 ($H_1H_2$). Second, we considered two RNAs with different structures (purine riboswitch with PDB code 2G9C and guanine riboswitch with PDB code 3RKF respectively) but annotated them with the same helical arrangement pattern, namely coaxial stacking helices 1 and 2 ($H_1H_2$). Note that this manually annotated $H_1H_2$ pattern is different from the $H_1H_3$ pattern that naturally occurs, and is also predicted by CHSalign_p, in the purine and guanine riboswitches.

In the first case, the score produced by CHSalign_u was very low (36.69), due to the different helical arrangements. This result shows the large conformational range of structural arrangements that the purine riboswitch can have, from naturally preferable arrangements ($H_1H_3$, as predicted by CHSalign_p) to unnatural arrangements ($H_1H_2$, as manually set by us). In the second case, CHSalign_u produced a high score of 179.68, which indicates the possibility that two different RNA structures can have very similar helical arrangements when we manually set these arrangements. Thus, CHSalign_u could help investigate the structural diversity of all possible helical arrangements, including natural or hypothetical conformations for two RNA 2D structures.

## Conclusions

We have presented a novel method (CHSalign) capable of producing an optimal alignment between two input RNA secondary (2D) structures with coaxial helical stacking (CHS), based on our previously developed Junction-Explorer [25] and RNAJAG [4]. The method is junction-aware, CHS-favored in the sense that it assigns a weight to the alignment of two RNA junctions with the same number of branches and the same coaxial helical stacking status while prohibiting the alignment of two junctions that do not have the same number of branches or the same coaxial helical stacking status. The method transforms each input RNA 2D structure to an ordered labeled tree, and employs dynamic programming techniques and a constrained tree matching algorithm to align the two input RNA 2D structures. CHSalign has two versions; CHSalign_u requires the user to manually annotate the coaxial stacking patterns in the input structures while CHSalign_p automatically predicts the coaxial stacking patterns in the input structures. Experimental results demonstrate that both versions outperform the existing alignment programs that do not take into account coaxial stacking configurations in the input RNA structures.

It has been observed that several functional RNA families such as tRNA, RNase P, and large ribosomal subunits have conserved structural features while having very diverse sequence patterns. RNA structure alignment tools such as CHSalign can help measure the structural similarity between these RNAs, even without sequence relevance in the RNAs. Similar RNA structural motifs are encountered on a variety of RNAs. While these motifs exist in different contexts, their functions are related. For instance, sarcin-ricin motifs often bind to proteins, and GNRA tetraloops act as receptors for RNA-RNA long-range interactions. Furthermore, examples of larger structure-function similarity are observed in the tRNA-like structure found in the transfer-messenger RNA (tmRNA), whose structure similarity with tRNA helps identify the functional role of tmRNAs to aid in translation via stalled ribosome rescue. Other tRNA-like structures found in viruses such as HIV and internal ribosome entry sites (IRES) mimic the 3D "L-shape" of tRNAs to take control of the host ribosome.

As our knowledge on RNA structure progresses, more sophisticated secondary structure alignment tools are required that allow for comparison of tertiary motifs such as coaxial stacking patterns. Indeed, experimental probing techniques such as RNA SHAPE chemistry, SAXS, NMR, and fluorescence resonance energy transfer (FRET), can often provide sufficient information to determine coaxial helical stacking [47,68,69]. Because the structure and function of RNA are highly interrelated, a tool that addresses coaxial stacking patterns can assist the comparison of structures with high functional relevance.

CHSalign is the first tool that can compute an RNA secondary structure alignment in the presence of coaxial helical stacking. When coaxial stacking configurations are available from experimental data such as FRET, NMR or SAXS data, the user can input such information to aid in the alignment. However, if no knowledge of coaxial stacking configurations is available, CHSalign can infer this information by employing Junction-Explorer [25], which predicts coaxial helical stacking with 81% accuracy.

Existing RNA secondary structure alignment tools [37,39] do not distinguish between structural elements such as helices, junctions and hairpin loops. However, each element type has its special property and function. In contrast, CHSalign only matches structural elements of the same type. Furthermore, the tool imposes a constraint that a junction of RNA1 can be aligned with a junction of RNA2 only if they have the same number of branches and the same coaxial helical stacking status. We also implemented an extension of CHSalign, which relaxes this constraint. This extension is able to align two junctions with different numbers of branches and simply requires that coaxially stacked helices be aligned with coaxially stacked helices when matching a p-way junction with a q-way junction for p different than q. The source code of both CHSalign and its extension can be downloaded from the web server site.

## Acknowledgments

We would like to acknowledge Drs. Bruce Shapiro and Kaizhong Zhang for helpful conversations. We also thank Dongrong Wen and Akhila Nagula for their contributions in the early stage of this work. J.W. acknowledges support of this work by the National Science Foundation Grant IIS-0707571. T.S. acknowledges support of this work by the National Institute of General Medical Sciences Grants GM100469 and GM081410. Computing resources, utilized by the NYU team, of the Computational Center for Nanotechnology Innovations and Empire State Development's Division of Science, Technology and Innovation [through National Science Foundation (NSF) Group Award TG-MCB080036N] and the New York Center for Computational Sciences at Stony Brook University/Brookhaven National Laboratory (supported by Department of Energy Grant DE-AC02-98CH10886 and the State of New York) are gratefully acknowledged.

# Figure Legends

**Fig 1. Transformation of an RNA 3D molecule into an ordered labeled tree.**

(A) The 3D crystal structure of the adenine riboswitch molecule (PDB code: 1Y26) obtained from the Protein Data Bank (PDB) and drawn by PyMOL. The first helix according to the 5′ to 3′ orientation is labeled by $H_1$ and highlighted in blue. The second helix is labeled by $H_2$ and highlighted in green. The third helix is labeled by $H_3$ and highlighted in red. The junction labeled by $J_1$ and hairpin loops labeled by $P_1$ and $P_2$ respectively are highlighted in light grey. $J_1$ is a multi-branch loop where the three helices $H_1$, $H_2$ and $H_3$ connect. $P_1$ and $P_2$ are hairpin loops connected to helices $H_2$ and $H_3$, respectively. (B) The corresponding secondary (2D) structure, obtained from RNAView. Each 2D structural element in (B) is highlighted as in (A). The yellow bar across $H_1$, $J_1$ and $H_3$ denotes a coaxial helical stacking $H_1H_3$ in the molecule 1Y26. (C) The ordered labeled tree, $T$, used to represent the 2D structure $R$ in (B). Each node of $T$ corresponds to a 2D structural element of $R$ where the octagon (squares, triangles respectively) in $T$ represents the junction (helices, hairpin loops respectively) in $R$.

**Fig 2. Illustration of an alignment between two RNA molecules.**

(A) The 3D crystal structure of the adenine riboswitch (PDB code: 1Y26) and its tree representation $T_1$. (B) The 3D crystal structure of the Alu domain of the mammalian signal recognition particle (SRP) (PDB code: 1E8O) and its tree representation $T_2$. (C) When matching $T_1[i]$ with $T_2[j]$, since $t_1[i]$ and $t_2[j]$ have different types where $t_1[i]$ is a junction and $t_2[j]$ is a helix, there are two subcases to be considered. Subcase 1 is illustrated in (i) where $t_2[j]$ is aligned to gaps and $T_1[i]$ is aligned with $T_2[j-1]$. Subcase 2 is illustrated in (ii) where $t_1[i]$ is aligned to gaps, and the subtree rooted at one of the children of $t_1[i]$ is aligned with $T_2[j]$. In this example, $t_1[i]$ has two children, $t_1[i_1]$ and $t_1[i_2]$. Thus, either the subtree rooted at $t_1[i_1]$, denoted by $T_1[i_1]$, is aligned with $T_2[j]$ as illustrated in (iia), or the subtree rooted at $t_1[i_2]$, denoted by $T_1[i_2]$, is aligned with $T_2[j]$ as illustrated in (iib).

**Fig 3. Comparison of the RMSD values obtained by CHSalign_u, CHSalign_p, RSmatch, RNAforester and FOLDALIGN.**

The RMSD values of CHSalign_u, CHSalign_p, RSmatch, RNAforester and FOLDALIGN are 1.78 Å, 1.83 Å, 4.41 Å, 6.13 Å and 8.26 Å, respectively. The proposed CHSalign method performs better than the existing alignment tools in terms of RMSD values.

**Fig 4. Comparison of the PR values obtained by CHSalign_u, CHSalign_p, RNAforester, SETTER, RSmatch and FOLDALIGN.**

The PR values, as defined in Equation (13), of CHSalign_u, CHSalign_p, RNAforester, SETTER, RSmatch and FOLDALIGN are 1, 0.85, 0.54, 0.42, 0.33 and 0.31, respectively. The proposed CHSalign method performs better than the existing alignment tools in terms of PR values.

**Fig 5. Illustration of the coaxial stacking patterns in the six riboswitches used to demonstrate the utility of our web server.**

(A) Artificial purine riboswitch (PDB code: 2G9C) with a three-way junction and a CHS motif of type $H_1H_3$ in the junction. (B) Artificial guanine riboswitch (PDB code:

3RKF) with a three-way junction and a CHS motif of type $H_1H_3$ in the junction. (C) *A. thaliana* TPP riboswitch (PDB code: 3D2G) with a three-way junction and a CHS motif of type $H_1H_2$ in the junction. (D) *E. coli* TPP riboswitch (PDB code: 2GDI) with a three-way junction and a CHS motif of type $H_1H_2$ in the junction. (E) *T. tengcongensis* SAM-I riboswitch (PDB code: 2GIS) with a four-way junction and a CHS motif of type $H_1H_4$, $H_2H_3$ in the junction. (F) *H. marismortui* SAM-I riboswitch (PDB code: 4B5R) with a four-way junction and a CHS motif of type $H_1H_4$, $H_2H_3$ in the junction.

# Tables

**Table 1. The 24 RNA full structures in Dataset1 selected from the Protein Data Bank (PDB) to evaluate the performance of the alignment methods studied in this paper.**

|    | PDB Code | Molecule Name | Length |
|----|----------|---------------|--------|
| 1  | 1E8O | Alu domain of the Signal recognition particle (7SL RNA) | 50 |
| 2  | 1L9A | Signal recognition particle RNA S domain | 126 |
| 3  | 1LNG | Signal recognition particle (7S.S RNA) | 97 |
| 4  | 1NBS | Ribonuclease P RNA | 119 |
| 5  | 1NKW | 23S ribosomal RNA | 2884 |
| 6  | 1NYI | Hammerhead ribozyme | 40 |
| 7  | 1S72 | 23S ribosomal RNA | 2876 |
| 8  | 1U6B | Group I intron | 222 |
| 9  | 1U8D | xpt-pbuX guanine riboswitch aptamer domain | 67 |
| 10 | 1UN6 | 5S ribosomal RNA | 122 |
| 11 | 1X8W | Tetrahymena ribozyme RNA (group I intron) | 968 |
| 12 | 1Y26 | Vibrio vulnificus A-riboswitch | 71 |
| 13 | 2A64 | Ribonuclease P RNA | 298 |
| 14 | 2AVY | 16S ribosomal RNA | 1530 |
| 15 | 2AW4 | 23S ribosomal RNA | 2958 |
| 16 | 2B57 | Guanine riboswitch | 65 |
| 17 | 2CKY | Thiamine pyrophosphate riboswitch | 154 |
| 18 | 2CZJ | Transfer-messenger RNA (tmRNA) | 248 |
| 19 | 2EES | Guanine riboswitch | 68 |
| 20 | 2GDI | TPP riboswitch | 80 |
| 21 | 2HOJ | THI-box riboswitch | 75 |
| 22 | 2J00 | 16S ribosomal RNA | 1687 |
| 23 | 2J01 | 23S ribosomal RNA | 2891 |
| 24 | 2QBZ | M-Box RNA, ykoK riboswitch aptamer | 153 |

**Table 2. The six riboswitches selected from the Protein Data Bank (PDB) to demonstrate the utility of our web server.**

|   | PDB Code | Molecule Name | Length | Junction | CHS |
|---|----------|---------------|--------|----------|-----|
| 1 | 2G9C | Artificial purine riboswitch | 68 | 3-way | $H_1H_3$ |
| 2 | 3RKF | Artificial guanine riboswitch | 68 | 3-way | $H_1H_3$ |
| 3 | 3D2G | *A. thaliana* TPP riboswitch | 77 | 3-way | $H_1H_2$ |
| 4 | 2GDI | *E. coli* TPP riboswitch | 80 | 3-way | $H_1H_2$ |
| 5 | 2GIS | *T. tengcongensis* SAM-I riboswitch | 95 | 4-way | $H_1H_4$ $H_2H_3$ |
| 6 | 4B5R | *H. marismortui* SAM-I riboswitch | 95 | 4-way | $H_1H_4$ $H_2H_3$ |

**Table 3. Results obtained by aligning seven pairs of riboswitches from Table 2.**

| Program | Molecule 1 | Molecule 2 | Alignment Score |
|---|---|---|---|
| CHSalign_p | 2GIS *T. tengcongensis* SAM-I riboswitch ($H_1H_4, H_2H_3$) | 4B5R *H. marismortui* SAM-I riboswitch ($H_1H_4, H_2H_3$) | 252.61 |
| CHSalign_p | 2G9C Artificial purine riboswitch ($H_1H_3$) | 3RKF Artificial guanine riboswitch ($H_1H_3$) | 179.68 |
| CHSalign_p | 2GDI *E. coli* TPP riboswitch ($H_1H_2$) | 3D2G *A. thaliana* TPP riboswitch ($H_1H_2$) | 191.06 |
| CHSalign_p | 2GIS *T. tengcongensis* SAM-I riboswitch ($H_1H_4, H_2H_3$) | 2G9C Artificial purine riboswitch ($H_1H_3$) | 20.40 |
| CHSalign_p | 2G9C Artificial purine riboswitch ($H_1H_3$) | 2GDI *E. coli* TPP riboswitch ($H_1H_2$) | 13.65 |
| CHSalign_u | 2G9C Artificial purine riboswitch ($H_1H_3$) | 2G9C Artificial purine riboswitch ($H_1H_2$) | 36.69 |
| CHSalign_u | 2G9C Artificial purine riboswitch ($H_1H_2$) | 3RKF Artificial guanine riboswitch ($H_1H_2$) | 179.68 |

# Supporting Information

**S1 File. Results obtained by aligning five pairs of riboswitches from Table 2 using CHSalign_p.**

For each pair of riboswitches, the input and output of the CHSalign_p program are displayed. The input includes two riboswitches in bpseq format. CHSalign_p invokes Junction-Explorer to predict coaxial helical stacking (CHS) motifs in the input molecules, and aligns the predicted structures. The output includes the CPU time spent in performing the alignment, the alignment score and alignment details.

**S2 File. Results obtained by aligning two pairs of riboswitches from Table 2 using CHSalign_u.**

For each pair of riboswitches, the input and output of the CHSalign_u program are displayed. The input includes two riboswitches in bpseq format along with CHS motifs annotated manually by the user. The output includes the CPU time spent in performing the alignment, the alignment score and alignment details.

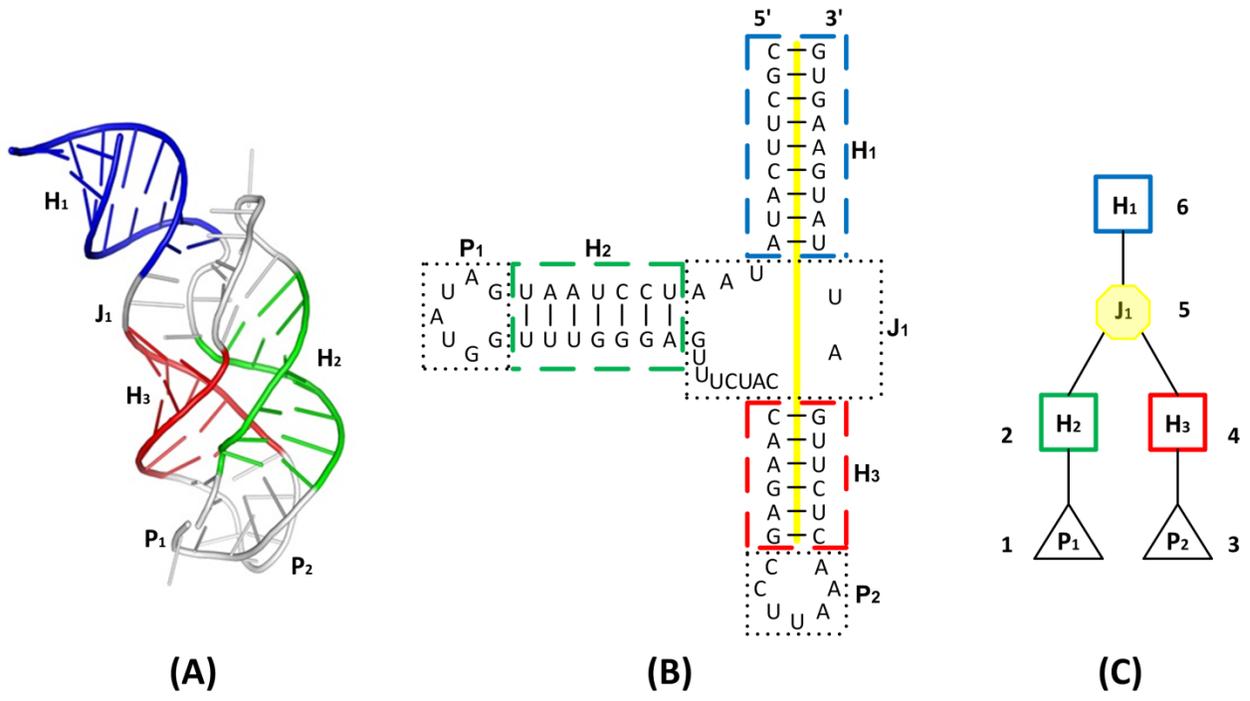

Figure 1

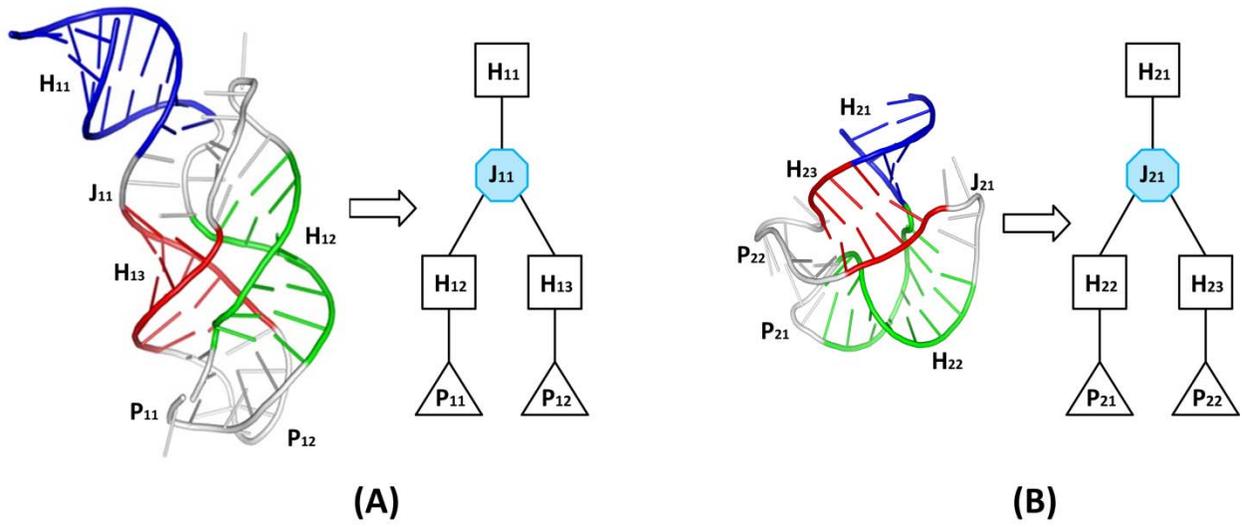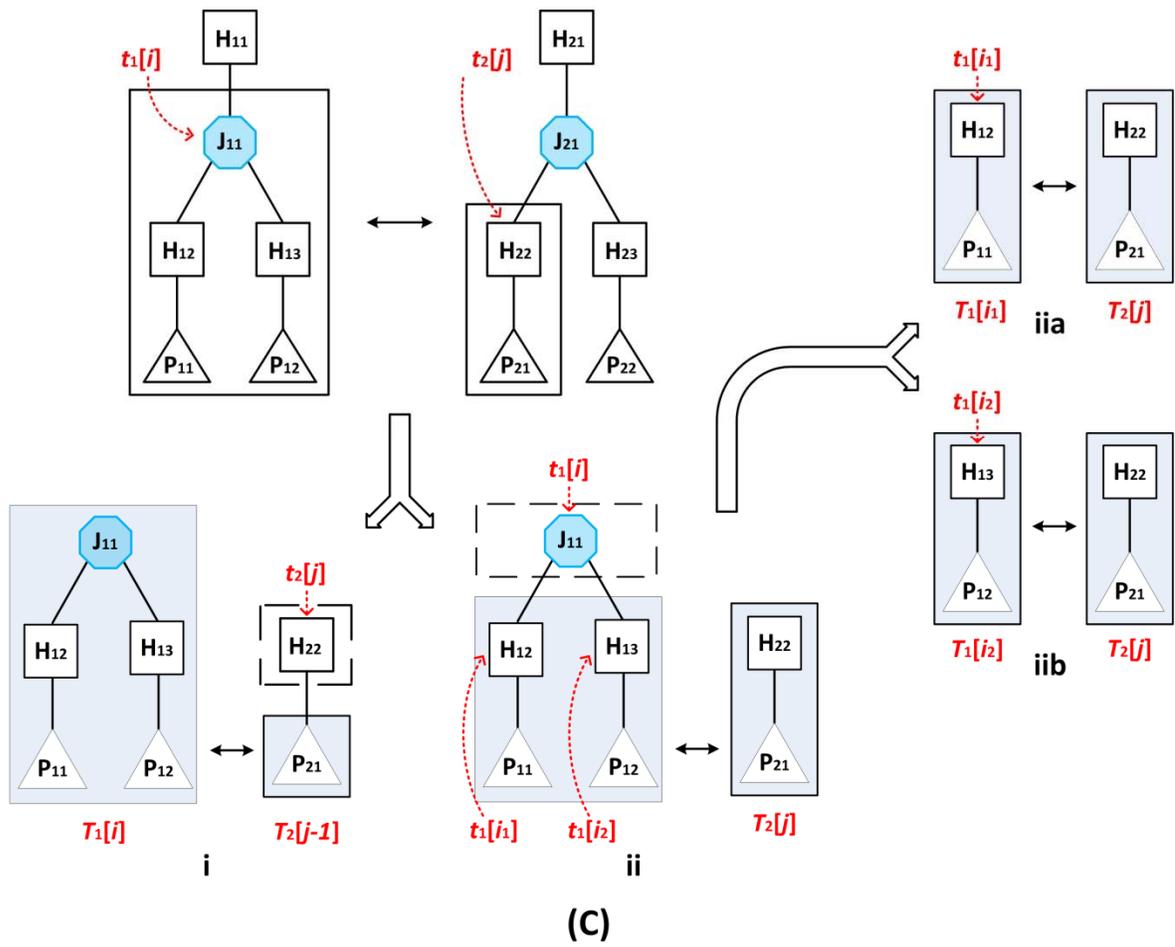

Figure 2

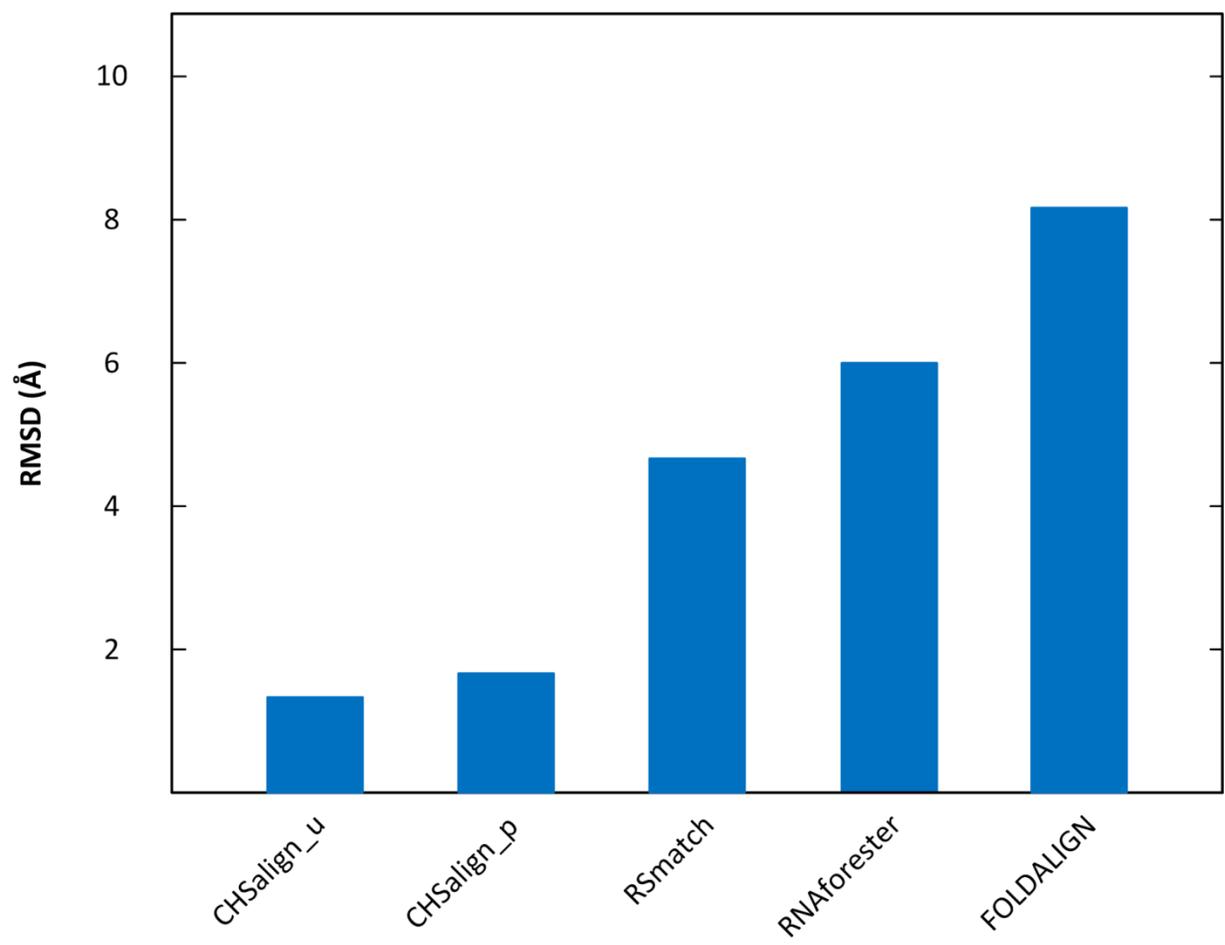

Figure 3

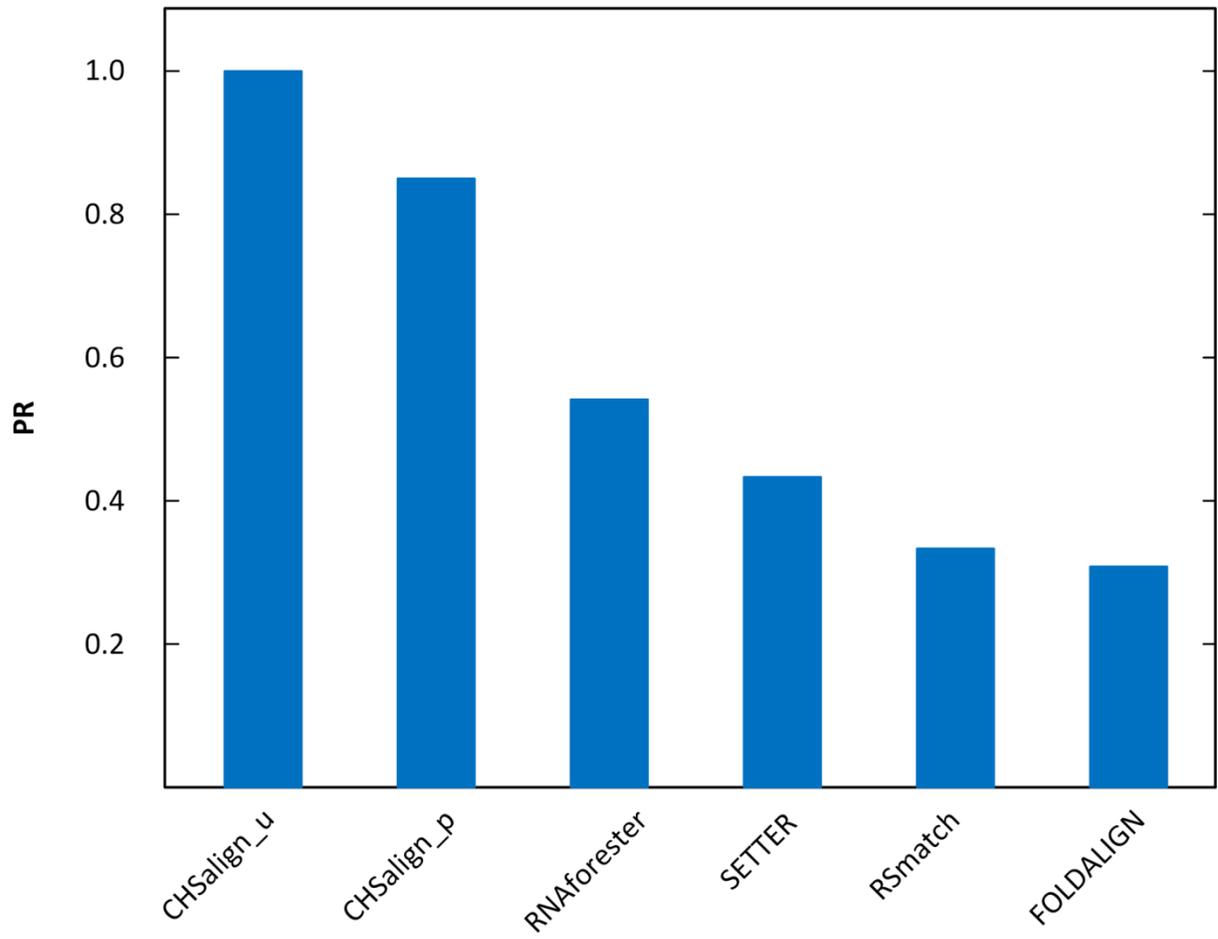

Figure 4

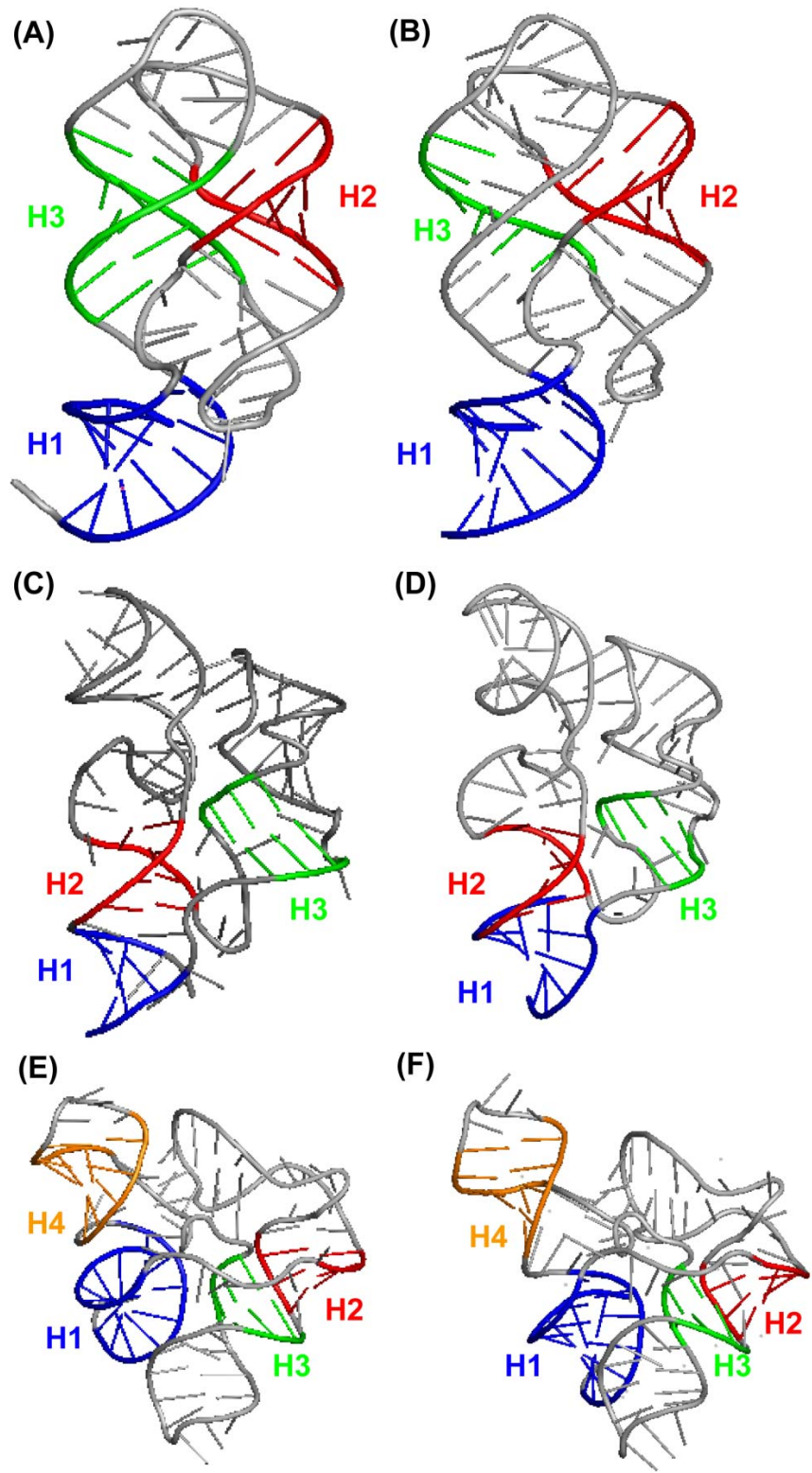

Figure 5